\documentclass[showpacs,preprintnumbers,aps]{revtex4}
\usepackage{graphicx}
\usepackage{bm}
\usepackage{color}
\usepackage{ulem}

\begin{document}

\title{Quantum dynamics of a binary mixture of BECs in a double well potential: an Holstein-Primakoff approach}
\author{Roberta Citro}
\email{citro@sa.infn.it}
\affiliation{Dipartimento di Fisica "E. R. Caianiello",
Universit\'{a} degli Studi di Salerno and CNR-SPIN, Unit\'{a} Operativa di Salerno, Via Ponte Don Melillo, 84084 Fisciano (SA),
Italy}
\author{Adele Naddeo}
\email{naddeo@sa.infn.it}
\affiliation{CNISM, Unit\'{a} di
Ricerca di Salerno and Dipartimento di Fisica "E. R. Caianiello",
Universit\'{a} degli Studi di Salerno, Via Ponte Don Melillo, 84084 Fisciano (SA),
Italy}
\author{Edmond Orignac}
\email{Edmond.Orignac@ens-lyon.fr}
\affiliation{Laboratoire de Physique, CNRS-UMR5672, {\'E}cole Normale Superieure de
Lyon, 46, All\'ee d'Italie, 69364 Lyon Cedex 07, France}

\date{\today}

\begin{abstract}
We study the quantum dynamics of a binary mixture of Bose-Einstein condensates (BEC) in a double-well potential
starting from a two-mode Bose-Hubbard Hamiltonian.  Focussing on the regime where the number of atoms is very large, a mapping onto a $SU(2)$ spin problem together with a Holstein-Primakoff transformation is performed. The quantum evolution of the number difference of bosons between the two wells is investigated for  different initial conditions, which range from the case of a small imbalance between the two wells to a coherent spin state. The results show an instability towards a phase-separation above a critical positive value of the interspecies interaction while the system evolves towards a coherent tunneling regime for negative interspecies interactions. A comparison with a semiclassical approach is discussed together with some implications on the experimental realization of phase separation with cold atoms.
\end{abstract}

 \maketitle


\section{Introduction}

Bose-Einstein condensates of dilute, weakly interacting gases offer a unique
possibility for exploring many-body dynamics, the role of quantum
fluctuations and in general macroscopic quantum coherence phenomena \cite
{leggett1}, thanks to a wide tunability of the interaction parameters.
Indeed several experimental strategies can be devised in order to pursue
this task, which range from the direct control via magnetic Feshbach
resonance techniques \cite{control1} to the transverse confinement in a
quasi one dimensional system \cite{control2} as a way to increase the
inter-atomic interaction. Finally, the introduction of an optical lattice
whose depth can be tuned allows one to decrease the kinetic term in the
Hamiltonian. Within the tight binding approximation such systems are
described by the Bose-Hubbard Hamiltonian, whose parameters are the hopping
frequency $\overline{E}_{J}$ between nearest neighbor lattice sites, the
onsite interaction strength $E_{c}$ and the total atoms number $N$. When the
ratio $\frac{E_{c}N}{\overline{E}_{J}}$ exceeds unity, a quantum phase
transition from a superfluid to a Mott insulator \cite{mott1} takes place
and the system enters a quantum regime characterized by strong correlations.
The simplest Hamiltonian of this kind that one can devise is the
Bose-Hubbard dimer \cite{dimer1}, which describes the physics of two weakly
coupled condensates. It can be mapped onto a $SU(2)$ spin problem and is
deeply related to the physics of Josephson junctions \cite{gati1}\cite
{ferrini1}\cite{leggett1}. Furthermore, if the mean field approximation is
considered one obtains the Gross-Pitaevskii theory which gives rise to a
variety of phenomena, ranging from Josephson oscillations \cite{jos1} to
macroscopic quantum self-trapping (MQST) \cite{smerzi1} and ac and dc
Josephson like effect \cite{smerzi2}, all experimentally observed in the
last decade \cite{exp1}\cite{exp2}\cite{exp3}.

More recently, after the experimental realization of two-species BECs \cite
{exp4}\cite{exp5}\cite{exp6}, the theoretical analysis on weakly coupled condensates has been successfully
extended to a binary mixture of BECs in a double well potential
\cite{mix0}\cite{mix1}\cite{mix2} \cite{mix3}\cite{mix4}\cite{mix4bis}. The semiclassical
regime in which the fluctuations around the mean values are small
has been deeply investigated and found to be described by two
coupled Gross-Pitaevskii equations. By means of a two-mode
approximation such equations can be cast in the form of four
coupled nonlinear ordinary differential equations for the
population imbalance and the relative phase of each species. The
solution results in a richer tunneling dynamics \cite{mix5}. In particular,
two different MQST states with broken symmetry have been found
\cite{mix3}, where the two species localize in the two different
wells giving rise to a phase separation or coexist in the same
well respectively.  Indeed, upon a variation of some parameters or
initial conditions, the phase-separated MQST states evolve towards
a symmetry-restoring phase where the two components swap places
between the two wells, so avoiding each other. Furthermore the
coherent dynamics of a two species BEC in a double well has been
analyzed as well focussing on the case where the two species are
two hyperfine states of the same alkali metal \cite{mix6}.

In a recent paper \cite{noi1} we studied the quantum behaviour of a binary
mixture of Bose-Einstein condensates (BEC) in a double-well potential
starting from a two-mode Bose-Hubbard Hamiltonian. We analyzed in detail the
small tunneling amplitude regime where number fluctuations are suppressed
and a Mott-insulator behaviour is established. Within this regime we
performed a perturbative calculation up to second order in the tunneling
amplitude and found the stationary states. In order to carry out analytical calculations we focused on
the symmetric case of equal nonlinear interaction and equal tunneling amplitude of the two species. Furthermore we restricted to the case in which the two species are equally populated and imposed the condition of equal population imbalance of the species $a$ and $b$ between the two wells. Then, the dynamics of the junction was investigated in correspondence of a
completely localized initial state. In order to avoid the above restrictions on the parameters range, here we focus
on the two-mode Bose-Hubbard Hamiltonian describing the two-species BEC ($a$ and $b$) in a double well when $N_{a},N_{b}\gg 1$, and perform a mapping onto a $SU(2)$ spin problem together with a Holstein-Primakoff transformation\cite{hp1}\cite{hp2}. As a result we obtain a Hamiltonian of two decoupled quantum harmonic oscillators, similar to that of Ref.\cite{hp3}, whose stationary states are readily found. The quantum evolution of the number difference of bosons between the two wells is investigated in detail in correspondence of a variety of initial conditions, which range from an initial state with small imbalance between the two species to a coherent spin state. The whole parameters space is explored by tuning the population, the tunneling amplitude and the nonlinear interaction for each species as well as the interspecies interaction in a wide range, from a symmetric to a strongly asymmetric case. Finally a detailed comparison with a semiclassical approach is given. Let us notice that Holstein-Primakoff transformation makes the system exactly solvable in the weakly interacting regime of interest in this work and that simplifies the study of the tunneling dynamics as well as the phase separation phenomenon. This is the main advantage of the approach chosen.

The paper is organized as follows.
In Section 2 we introduce the model Hamiltonian we study within the two mode approximation. A Holstein-Primakoff transformation is performed and the semiclassical limit is taken followed by a decoupling of the bosonic degrees of freedom for each species. As a result the Hamiltonian can be rephrased in terms of two independent harmonic oscillators, whose stationary states are derived in Section 3. In Section 4 the quantum dynamics of the system is discussed in correspondence of two different initial conditions: small imbalance between the two wells and coherent states. A wide range of values of interspecies interaction is explored and the crossover to an unstable regime with phase separation is found. In Section 5 the classical equations of motion are derived and a comparison with the quantum counterpart is carried out. Finally some conclusions and perspectives of this work are briefly outlined.

\section{The model}

A binary mixture of Bose-Einstein condensates \cite{mix1}\cite{mix3} loaded
in a double-well potential is described by the
Hamiltonian $H=H_{a}+H_{b}+H_{ab}$ where:
\begin{equation}
H_{i}=\int d\overrightarrow{r}\left( -\frac{\hbar ^{2}}{2m_{i}}\psi
_{i}^{+}\nabla ^{2}\psi _{i}+\psi _{i}^{+}V_{i}\left( \overrightarrow{r}%
\right) \psi _{i}\right) +\frac{g_{ii}}{2}\int d\overrightarrow{r}\psi
_{i}^{+}\psi _{i}^{+}\psi _{i}\psi _{i};\text{ \ \ \ }i=a,b  \label{m6}
\end{equation}
\begin{equation}
H_{ab}=g_{ab}\int d\overrightarrow{r}\psi _{a}^{+}\psi _{b}^{+}\psi _{a}\psi
_{b}.  \label{m7}
\end{equation}
Here $g_{ii}=\frac{4\pi \hbar ^{2}a_{ii}}{m_{i}}$ is the intraspecies coupling constants, being  $m_{i}$ the atomic mass
and $a_{ii}$ the $s$-wave scattering lengths;
$g_{ab}=\frac{2\pi \hbar ^{2}a_{ab}}{m_{ab}}$ is the interspecies coupling
constant, where $m_{ab}=\frac{m_{a}m_{b}}{m_{a}+m_{b}}$ is the reduced mass;
$V_{i}\left( \overrightarrow{r}\right) $ is the double well trapping
potential and, in the following, we assume $V_{a}\left( \overrightarrow{r}%
\right) =V_{b}\left( \overrightarrow{r}\right) =V\left( \overrightarrow{r}%
\right) $; $\psi _{i}^{+}\left( \overrightarrow{r}\right) ,$ $\psi
_{i}\left( \overrightarrow{r}\right) $, $i=a,b$ are the bosonic creation and
annihilation operators for the two species, which satisfy the commutation
rules:
\begin{eqnarray}
\left[ \psi _{i}\left( \overrightarrow{r}\right) ,\psi _{j}\left(
\overrightarrow{r}^{\prime }\right) \right] &=&\left[ \psi _{i}^{+}\left(
\overrightarrow{r}\right) ,\psi _{j}^{+}\left( \overrightarrow{r}^{\prime
}\right) \right] =0,  \label{m8} \\
\left[ \psi _{i}\left( \overrightarrow{r}\right) ,\psi _{j}^{+}\left(
\overrightarrow{r}^{\prime }\right) \right] &=&\delta _{ij}\delta \left(
\overrightarrow{r}-\overrightarrow{r}^{\prime }\right) ,\text{ \ \ \ \ }%
i,j=a,b,  \label{m9}
\end{eqnarray}
and the normalization conditions:
\begin{equation}
\int d\overrightarrow{r}\left| \psi _{i}\left( \overrightarrow{r}\right)
\right| ^{2}=N_{i};\text{ \ \ \ }i=a,b,  \label{m10}
\end{equation}
$N_{i}$, $i=a,b$ being the number of atoms of species $a$ and $b$
respectively. The total number of atoms of the mixture is $N=N_{a}+N_{b}$.

A weak link between the two wells produces a small energy splitting
between the mean-field ground state and the first excited state of the
double well potential and that allows to reduce the dimension of the
Hilbert space of the initial many-body problem. Indeed for low energy
excitations and low temperatures it is possible to consider only such two
states and neglect the contribution from the higher ones, the so called
two-mode approximation \cite{milburn} \cite{smerzi1}\cite{ananikian1}.
In this approximation the Hamiltonian (\ref{m6}) can be written in terms of the
the annihilation operators, $a_{L}=\frac{1}{\sqrt{2}}\left(
a_{g}+a_{e}\right) $, $a_{R}=\frac{1}{\sqrt{2}}\left( a_{g}-a_{e}\right) $
and $b_{L}=\frac{1}{\sqrt{2}}\left( b_{g}+b_{e}\right) $, $b_{R}=\frac{1}{%
\sqrt{2}}\left( b_{g}-b_{e}\right) $ and the corresponding creation operators, where $a_{g},a_{e}$ and $b_{g},b_{e}$ are the annihilation operators of a particle in the ground and in the first excited state.

When introducing the angular momentum operators:
\begin{equation}
\begin{array}{ccc}
J_{x}^{a}=\frac{1}{2}\left( a_{R}^{+}a_{L}+a_{L}^{+}a_{R}\right) , &
J_{y}^{a}=\frac{i}{2}\left( a_{R}^{+}a_{L}-a_{L}^{+}a_{R}\right) , &
J_{z}^{a}=\frac{1}{2}\left( a_{R}^{+}a_{R}-a_{L}^{+}a_{L}\right) , \\
J_{x}^{b}=\frac{1}{2}\left( b_{R}^{+}b_{L}+b_{L}^{+}b_{R}\right) , &
J_{y}^{b}=\frac{i}{2}\left( b_{R}^{+}b_{L}-b_{L}^{+}b_{R}\right) & J_{z}^{b}=%
\frac{1}{2}\left( b_{R}^{+}b_{R}-b_{L}^{+}b_{L}\right) ,
\end{array}
\label{m23}
\end{equation}
where the operators $J_{i}^{a}$, $J_{i}^{b}$, $i=x,y,z$, obey to the usual
angular momentum algebra together with the relation:
\begin{equation}
\begin{array}{cc}
\left( J^{a}\right) ^{2}=\frac{N_{a}}{2}\left( \frac{N_{a}}{2}+1\right) , &
\left( J^{b}\right) ^{2}=\frac{N_{b}}{2}\left( \frac{N_{b}}{2}+1\right) ,
\end{array}
\label{m23a}
\end{equation}
the Hamiltonian of the double species Bose-Josephson junction can be written in the form:
\begin{eqnarray}
H &=&\frac{1}{2}\Lambda _{a}\left( J_{z}^{a}\right)
^{2}-K_{a}J_{x}^{a}+C_{a}\left( J_{x}^{a}\right) ^{2}+\frac{1}{2}\Lambda
_{b}\left( J_{z}^{b}\right) ^{2}-K_{b}J_{x}^{b}+C_{b}\left( J_{x}^{b}\right)
^{2}+  \nonumber \\
&&+\Lambda _{ab}J_{z}^{a}J_{z}^{b}-D_{ab}J_{x}^{a}J_{x}^{b}.  \label{m26}
\end{eqnarray}
where $K_{a,b}$ are the tunneling amplitudes between the two wells, $\Lambda_{a,b},\Lambda _{ab}$ are the intra- and interspecies interactions respectively,
while the terms $C_a$ and $D_{a,b}$ describe two-particle processes
\cite{noi1}. The form~(\ref{m26}) was previously discussed in the
  classical limit in \cite{mix3}, where it was shown to be lead to
  equations of motion equivalent to the Gross-Pitaevskii equations.
For $\Lambda_{ab}=D_{ab}=0$ in Eq.~(\ref{m26}),  the Hamiltonian
  reduces to a sum of two Lipkin-Meshkov-Glick (LMG) model \cite{lipkin1,lipkin2}
  Hamiltonian, one for each species. For $\Lambda_{ab}\ne 0$ or
$D_{ab} \ne 0$, the two LMG models are coupled.
Within the experimental parameters range it is possible to show that $%
C_{i}\ll \Lambda _{i},K_{i}$, $i=a,b$, and $D_{ab}\ll \Lambda _{ab}$ \cite{noi1}\cite
{gati1}\cite{mix3}, then in the following we put $C_{a}=C_{b}=0$ and $%
D_{ab}=0$, which corresponds to neglecting the spatial overlap integrals
between the localized modes in the two wells. In this way the binary mixture
of BECs within two-mode approximation maps to a two Ising spins model in a
transverse magnetic field, whose Hamiltonian is:
\begin{equation}
H=\frac{1}{2}\Lambda _{a}\left( J_{z}^{a}\right) ^{2}-K_{a}J_{x}^{a}+\frac{1%
}{2}\Lambda _{b}\left( J_{z}^{b}\right) ^{2}-K_{b}J_{x}^{b}+\Lambda
_{ab}J_{z}^{a}J_{z}^{b}.  \label{m27}
\end{equation}
Let us briefly discuss the symmetries of the
Hamiltonian~(\ref{m27}). First, when $\Lambda_{ab}=0$, the Hamiltonian
decouples into $H=H_a+H_b$, and $e^{i\pi J_x^{\nu}} H_{\nu'} e^{-i\pi
  J_x^{\nu}}=H_{\nu'}$ for $\nu,\nu' \in \{a,b\}$. Therefore, the
eigenstates of $H$ can be sought in the form of eigenstates of
$e^{i\pi J_x^{a}}$ and $e^{i\pi J_x^{b}}$. Since $e^{2i\pi
  J_x^\nu}=e^{i \pi N_\nu}$, these eigenvalues are $\pm 1$ when
$N_\nu$ is even, and $\pm i$ when $N_\nu$ is odd. So the Hilbert space
breaks down into four sectors indexed by the eigenvalues of   $e^{i\pi
  J_x^{a}}$ and $e^{i\pi J_x^{b}}$. Then, turning on $ \Lambda_{ab}\ne
0$, the only remaining symmetry is $ e^{i\pi (J_x^{a}+J_x^{b})} H_{\nu'} e^{-i\pi
  (J_x^{a}+J_x^{b}) }$, so that only two independent sectors
remain. These sectors are formed by the combination in pairs of the four
sectors obtained for $\Lambda_{ab}=0$.

Let us now make the rotation:
\begin{equation}
\begin{array}{cc}
\begin{array}{c}
J_{z}^{i}\rightarrow -J_{x}^{i} \\
J_{x}^{i}\rightarrow J_{z}^{i}
\end{array}
, & i=a,b.
\end{array}
\label{m28}
\end{equation}
To proceed we perform the Holstein-Primakoff transformation \cite{hp1,hp2,hp3} in order to map the angular momentum operators into bosonic ones and focus on the regime with large number of atoms $N_{a},N_{b}\gg 1$ and weak scattering strengths $K_{a(b)}\gg \Lambda _{a}, \Lambda _{b}, \Lambda _{ab}$:
\begin{equation}
\begin{array}{c}
J_{z}^{a}=J^{a}-a^{\dagger}a \\
J_{+}^{a}= \sqrt{2J^{a}-a^\dagger a}a \\
J_{-}^{a}= a^\dagger \sqrt{2J^{a} -a^\dagger a }
\end{array}
,
\begin{array}{c}
J_{z}^{b}=J^{b}-b^{\dagger}b \\
J_{+}^{b}= \sqrt{2J^{b}-b^{\dagger}b}b \\
J_{-}^{b}= b^\dagger \sqrt{2J^{b}-b^{\dagger}b}
\end{array}
,  \label{m29}
\end{equation}
where $J_{\pm}^{i}=J_{x}^{i}\pm iJ_{y}^{i},\; i=a,b$, $J^i=N_i/2\;
  i=a,b$, thus leading to the Hamiltonian:
\begin{eqnarray}
  \label{eq:holstein-full}
  H&=&\frac{\Lambda_a}{8} \left[ 2 J^a(J^a+1) - 2 (J^a - a^\dagger a)^2
    + \sqrt{(2 J^a -a^\dagger a)(2J^a -1 - a^\dagger a)} a^2 +
    (a^\dagger)^2 \sqrt{(2 J^a -a^\dagger a)(2J^a -1 - a^\dagger a)}
  \right] \nonumber \\
&&+\frac{\Lambda_b}{8} \left[ 2 J^b(J^b+1) - 2 (J^b - b^\dagger b)^2
    + \sqrt{(2 J^b -b^\dagger b)(2J^b -1 - b^\dagger b)} b^2 +
    (b^\dagger)^2 \sqrt{(2 J^a -b^\dagger b)(2J^a -1 - b^\dagger b)}
  \right]  \nonumber \\
&& + \frac{\Lambda_{ab}} 4 \left[   \sqrt{(2 J^a -a^\dagger a)(2 J^b
    -b^\dagger b)} a b + b^\dagger a^\dagger  \sqrt{(2 J^a -a^\dagger a)(2 J^b
    -b^\dagger b)} \right. \nonumber \\
&& \left. +  \sqrt{2 J^b -b^\dagger b} a^\dagger b \sqrt{2
    J^a -a^\dagger a} +  \sqrt{2
    J^a -a^\dagger a} b^\dagger a  \sqrt{2
    J^a -a^\dagger a} \right]\nonumber \\
&& + K_a (a^\dagger a - J^a) + K_b ( b^\dagger b - J^b).
\end{eqnarray}
Here $a$ and $b$ are boson annihilation operators for each species. In this representation, the operators $e^{i\pi J_z^{\nu}}$ are equal to
$e^{i\pi (J^\nu -\nu^\dagger \nu)}$ and their action is simply $\nu
\to -\nu$. For $\Lambda_{ab}=0$, the Hilbert space of
(\ref{eq:holstein-full}) thus breaks into four different sectors, according to
the parity of $a^\dagger a$ and $b^\dagger b$, while for
$\Lambda_{ab}\ne 0$, it breaks into two different sectors depending on
the parity of $a^\dagger a + b^\dagger b$. The physical Hilbert space
is restricted to $0\le a^\dagger a \le N_a$ and $0 \le b^\dagger b
\le N_b$.

Since we are considering a large number of atoms, we have $J^{a},J^{b}\gg 1$, while the condition $K_{a(b)}\gg \Lambda _{a}, \Lambda _{b}, \Lambda _{ab}$ implies $\langle a^{+}a \rangle \ll 2J^a$ and $\langle b^{+}b \rangle \ll 2J^b$. Under these assumptions one can use the linearized Holstein-Primakoff transformation \cite{hp1} (i.e. $J_{z}^{s}=J^{s}-s^{\dagger}s, J_{+}^{s}= \sqrt{2J^{s}}s, J_{-}^{s}= s^\dagger \sqrt{2J^{s}}$ with $s=a,b$) and derive the effective Hamiltonian:
\begin{eqnarray}
H &=&\Lambda _{a}J^{a}\left( \frac{a+a^{+}}{2}\right) \left( \frac{a+a^{+}}{2%
}\right) +\Lambda _{b}J^{b}\left( \frac{b+b^{+}}{2}\right) \left( \frac{%
b+b^{+}}{2}\right) +  \nonumber \\
&&2\Lambda _{ab}\sqrt{J^{a}J^{b}}\left( \frac{a+a^{+}}{2}\right) \left(
\frac{b+b^{+}}{2}\right) -K_{a}J^{a}-K_{b}J^{b}+K_{a}a^{+}a+K_{b}b^{+}b.
\label{m30}
\end{eqnarray}
In order to decouple the degrees of freedom of
each bosonic species let us introduce the following harmonic oscillator
coordinates and momenta, $q_{i}$, $p_{i}$, $i=a,b$:
\begin{equation}
\begin{array}{cc}
q_{a}=\frac{1}{\sqrt{2}}\left( a+a^{+}\right) , & q_{b}=\frac{1}{\sqrt{2}}%
\left( b+b^{+}\right) \\
p_{a}=\frac{-i}{\sqrt{2}}\left( a-a^{+}\right) & p_{b}=\frac{-i}{\sqrt{2}}%
\left( b-b^{+}\right)
\end{array}
,  \label{m31}
\end{equation}
which satisfy the usual commutation rules $\left[ q_{i},p_{j}\right]
=i\delta _{ij}$, $i,j=a,b$. Then, by defining:
\begin{equation}
\begin{array}{cc}
Q_{a}=\frac{q_{a}}{\sqrt{K_{a}}}, & Q_{b}=\frac{q_{b}}{\sqrt{K_{b}}}, \\
P_{a}=\sqrt{K_{a}}p_{a}, & P_{b}=\sqrt{K_{b}}p_{b},
\end{array}
\label{m32}
\end{equation}
(where $\left[ Q_{i},P_{j}\right] =i\delta _{ij}$, $i,j=a,b$) and, by
dropping constant terms, Eq. (\ref{m30}) can be written in a matrix form as \cite{hp3}:
\begin{equation}
\widehat{H}_{2BJJ}\simeq \frac{1}{2}\left[ \hat{Q}^{T}\widehat{\omega}^{2}\hat{Q}+\hat{P}^{T}\hat{P}\right] ,
\label{m33}
\end{equation}
where
\begin{equation}
\widehat{\omega}^2=
\left(\begin{array}{cc}
  \omega_a^2 & \omega_{ab} \\
  \omega_{ab} & \omega_{b}^2
\end{array}
\right)
\end{equation}
and $\hat{Q}^T=(Q_a,Q_b)$, $\hat{P}^T=(P_a,P_b)$ (the symbol $\cdot^T$ stands for the transpose);
$\omega _{i}^{2}=\Lambda _{i}J^{i}K_{i}+K_{i}^{2}$, and $\omega _{ab}=\Lambda _{ab}%
\sqrt{J^{a}J^{b}K_{a}K_{b}}$.

A straightforward diagonalization gives the Hamiltonian:
\begin{equation}
H_{2BJJ}\simeq \frac{1}{2}\left[ \omega _{1}^{2}Q_{1}^{2}+P_{1}^{2}+\omega
_{2}^{2}Q_{2}^{2}+P_{2}^{2}\right] ,\label{m34}
\end{equation}
where, defining $\Delta_{ab}=\sqrt{\left(\omega_{a}^{2}-\omega _{b}^{2}\right) ^{2}+4\omega _{ab}^{2}}$,
\begin{equation}
\begin{array}{cc}
\omega_{1}^{2}=\frac{\omega _{a}^{2}+\omega _{b}^{2}-\Delta_{ab}}{2}, & \omega_{2}^{2}=\frac{\omega _{a}^{2}+\omega _{b}^{2}+\Delta_{ab}}{2}
\end{array}
,  \label{m35}
\end{equation}
\begin{equation}
\begin{array}{cc}
Q_{1}=\frac{\left\{ 2\omega _{ab}Q_{b}-\left[ \left( \omega _{b}^{2}-\omega
_{a}^{2}\right) +\Delta_{ab}\right] Q_{a}\right\} }{\sqrt{4\omega _{ab}^{2}+\left[
\left( \omega _{b}^{2}-\omega _{a}^{2}\right) +\Delta_{ab}\right]^{2}}}, &
Q_{2}=\frac{\left\{ 2\omega _{ab}Q_{b}-\left[ \left( \omega _{b}^{2}-\omega
_{a}^{2}\right) -\Delta_{ab}\right] Q_{a}\right\} }{\sqrt{4\omega _{ab}^{2}+\left[
\left( \omega _{b}^{2}-\omega _{a}^{2}\right)-\Delta_{ab}\right]^{2}}},
\end{array}
\label{m36}
\end{equation}
\begin{equation}
\begin{array}{cc}
P_{1}=\frac{\left\{ 2\omega _{ab}P_{b}-\left[\left( \omega _{b}^{2}-\omega
_{a}^{2}\right) +\Delta_{ab}\right] P_{a}\right\}}{\sqrt{4\omega _{ab}^{2}+
\left[\left( \omega _{b}^{2}-\omega _{a}^{2}\right) +\Delta_{ab}\right]^{2}}}, &
P_{2}=\frac{\left\{ 2\omega _{ab}P_{b}-\left[ \left( \omega _{b}^{2}-\omega
_{a}^{2}\right) -\Delta_{ab}\right] P_{a}\right\} }{\sqrt{4\omega _{ab}^{2}+\left[
\left( \omega _{b}^{2}-\omega _{a}^{2}\right) -\Delta_{ab}\right] ^{2}}}.
\end{array}
\label{m37}
\end{equation}
The operators $Q_{1},P_{1}$ and $Q_{2},P_{2}$ can be viewed as
position and momentum operators of two distinct fictitious particles,
associated with the modes $1$ and $2$, i.e. the Hamiltonian
(\ref{m34}) is that of two harmonic oscillators.

The eigenvalues $\omega_{1,2}$ up to order $K_i^2$ obtained within the
Holstein-Primakoff approach coincide with the zero mode frequencies of small amplitude oscillations obtained by the semiclassical approach based on the Gross-Pitaevskii equations for the two condensate wave functions in Ref. \cite{mix1} (see Eq. (26)) and in Ref. \cite{mix3} (see Equation at the beginning of Section IV) for the case of equally populated species.
When $\omega_1^2$ vanishes, a phase separation takes place,
resulting in a MQST state.

Indeed, from Eqs. (\ref{m35}) the  stability condition is:
\begin{equation}
\left| \Lambda_{ab} \right|< \sqrt{\left(\Lambda_{a}+\frac{K_{a}}{J^a}\right) \left(\Lambda_{b}+\frac{K_{b}}{J^b}\right)}=\Lambda^c_{ab},\label{stab1}
\end{equation}
where $\Lambda^c_{ab}$ is the critical value of the interspecies
interaction which sets the onset of phase separation regime. Such a
condition agrees the one given in Ref. \cite{mix3} (see Equation (10) in Section IV) and reduces to:
\begin{equation}
\left| \Lambda_{ab} \right|< \sqrt{\Lambda_{a}\Lambda_{b}},\label{stab2}
\end{equation}
when the limit $J^a,J^b \rightarrow \infty $ is taken.

In the symmetric case $\Lambda_{a}=\Lambda_{b}=\Lambda$, $K_{a}=K_{b}=K$, $N_{a}=N_{b}=\frac{N}{2}$ we get $\omega _{a}^{2}=\omega _{b}^{2}=\omega^{2}$ where $\omega^{2}=\Lambda \frac{N}{2}K +K^{2}$, and $\omega _{ab}=\Lambda _{ab}\frac{N}{2} K$. As a consequence $\Delta_{ab}=2\omega _{ab}$ and the eigenvalues (\ref{m35}) simplify as:
\begin{equation}
\begin{array}{cc}
\omega_{1}^{2}=\omega^2 -\omega_{ab}, & \omega_{2}^{2}=\omega^2 +\omega_{ab}
\end{array}
,  \label{m35new}
\end{equation}
which result in the stability condition:
\begin{equation}
\left| \Lambda_{ab} \right|< \Lambda +2 \frac{K}{N}=\Lambda^c_{ab}.\label{stab2}
\end{equation}

In the next Sections we will derive the analytical expressions for the stationary states and discuss the corresponding quantum dynamics of the system.

\section{Stationary states}

Since the Hamiltonian (\ref{m34}) is that of two independent particles $H=H_1+H_2$,
the corresponding Hilbert space is simply given by the tensor product $%
\mathcal{E}_{a}\otimes \mathcal{E}_{b}\equiv \mathcal{E}_{1}\otimes \mathcal{%
E}_{2}$ and we can find a basis of
eigenvectors for $H_{2BJJ}$ in the following form: $\left| \varphi
\right\rangle =\left| \varphi ^{1}\right\rangle \left| \varphi
^{2}\right\rangle $, where $\left| \varphi ^{1}\right\rangle $ and $\left|
\varphi ^{2}\right\rangle $ are eigenvectors of $H_{1}$ and $H_{2}$ within $%
\mathcal{E}_{1}$ and $\mathcal{E}_{2}$. Since $H_{1}$ and $H_{2}$ are simply
harmonic oscillator Hamiltonians, we could define two pairs of creation
and annihilation operators, one for each mode, as follows:
\begin{equation}
\begin{array}{cc}
a_{i}^{+}=\frac{1}{\sqrt{2}}\left[ \sqrt{\frac{\omega _{i}}{\hbar }}Q_{i}-i%
\frac{P_{i}}{\sqrt{\omega _{i}\hbar }}\right],
\end{array}
\label{m38}
\end{equation}
\begin{equation}
\begin{array}{cc}
a_{i}=\frac{1}{\sqrt{2}}%
\left[ \sqrt{\frac{\omega _{i}}{\hbar }}Q_{i}+i\frac{P_{i}}{\sqrt{\omega
_{i}\hbar }}\right] ,
\end{array}
\label{m39}
\end{equation}
being $i=1,2$. Now, if we define the ground states of $H_{1}$ and $H_{2}$ as $\left|
\varphi _{0}^{1}\right\rangle $ and $\left| \varphi _{0}^{2}\right\rangle $,
we easily obtain eigenvalues and eigenvectors within these two subspaces as:
\begin{equation}
\begin{array}{cc}
E_{n}^{1}=\left( n+\frac{1}{2}\right) \hbar \omega _{1}, & \left| \varphi
_{n}^{1}\right\rangle =\frac{1}{\sqrt{n!}}\left( a_{1}^{+}\right) ^{n}\left|
\varphi _{0}^{1}\right\rangle ,
\end{array}
\label{m40}
\end{equation}
\begin{equation}
\begin{array}{cc}
E_{p}^{2}=\left( p+\frac{1}{2}\right) \hbar \omega _{2}, & \left| \varphi
_{p}^{2}\right\rangle =\frac{1}{\sqrt{p!}}\left( a_{2}^{+}\right) ^{p}\left|
\varphi _{0}^{2}\right\rangle .
\end{array}
\label{m41}
\end{equation}
So the stationary states of the full Hamiltonian (\ref{m34}) are:
\begin{equation}
\left| \varphi _{n,p}\right\rangle =\left| \varphi _{n}^{1}\right\rangle
\left| \varphi _{p}^{2}\right\rangle =\frac{1}{\sqrt{n!p!}}\left(
a_{1}^{+}\right) ^{n}\left( a_{2}^{+}\right) ^{p}\left| \varphi
_{0,0}\right\rangle ,  \label{m42}
\end{equation}
and the corresponding energies are:
\begin{equation}
E_{n,p}=E_{n}^{1}+E_{p}^{2}=\left( n+\frac{1}{2}\right) \hbar \omega
_{1}+\left( p+\frac{1}{2}\right) \hbar \omega _{2}.  \label{m43}
\end{equation}
We note that since the Hamiltonian~(\ref{m30}) preserved the
  original parity symmetry of the original
  Hamiltonian~(\ref{eq:holstein-full}), its eigenstates could also be
  classify according to their parity under $a\to -a$ and $b\to -b$.
Since $a_1$ and $a_2$ are linear combinations of $a,b$, the
eigenstates can also be classified by their parity under $a_{1,2}\to
-a_{1,2}$. Using (\ref{m42}), it is then clear that the even
eigenstates are those with $n+p$ even and the odd eigenstates the ones
with $n+p$ odd. So we can define the parity of a state as
$(-1)^{n+p}$.

We stress that this spectrum is not unbounded because an infinite number of unphysical high energy states
  have been added. Thus a constraint has to be included in order to satisfy the conditions $\langle a^\dagger a \rangle \ll 2J^a$, $\langle
b^\dagger b \rangle \ll 2J^b$. Solving these constraints will give
limits to the value of $n$ and $p$ and we will recover a finite dimensional
Hilbert space. Let us notice that, through the repeated action of the operators $a_{1}^{+}$
and $a_{2}^{+}$, we can obtain stationary states of the system with a
given number of quanta in each mode. The action of $a_{1}^{+}$, $a_{1}$,
$a_{2}^{+}$, $a_{2}$ on the stationary states $\left| \varphi
_{n,p}\right\rangle $ is as follows:
\begin{eqnarray}
a_{1}^{+}\left| \varphi _{n,p}\right\rangle =\sqrt{n+1}\left| \varphi
_{n+1,p}\right\rangle &,&  a_{1}\left| \varphi _{n,p}\right\rangle =\sqrt{n}\left| \varphi
_{n-1,p}\right\rangle  \label{m44b} \\
a_{2}^{+}\left| \varphi _{n,p}\right\rangle =\sqrt{p+1}\left| \varphi
_{n,p+1}\right\rangle &,& a_{2}\left| \varphi _{n,p}\right\rangle =\sqrt{p}\left| \varphi
_{n,p-1}\right\rangle . \label{m44d}
\end{eqnarray}
Generically,  $\omega _{1}$ and $\omega _{2}$ are incommensurate with
each other and there are no degenerate levels since there do not exist two
different pairs of integers $\left\{ n,p\right\} $ and $\left\{ n^{^{\prime
}},p^{^{\prime }}\right\} $ such that $n\omega _{1}+p\omega _{2}=n^{^{\prime
}}\omega _{1}+p^{^{\prime }}\omega _{2}$. Such degeneracy may exist in
the non-generic case where the ratio $\frac{\omega _{1}}{\omega _{2}}
$ is a rational number. In the presence of degeneracy, the non-linear
terms that we have neglected can lift the degeneracy, unless the
states have different parity.

\section{Quantum dynamics}

We are interested in the time evolution of the mean
values of the observables $J_{x}^{a}$, $J_{x}^{b}$%
, that is the population imbalance between the left and right well of the
potential of each bosonic species. In order to carry out such a program and to impose the correct
initial conditions it is much more convenient to start from the Heisenberg
equations of motion for the observables $Q_{1}$, $Q_{2}$, $P_{1}$, $P_{2}$:
\begin{eqnarray}
\frac{d}{dt}\left\langle Q_{i}\right\rangle &=&\frac{1}{i\hbar }\left\langle %
\left[ Q_{i},H_{2BJJ}\right] \right\rangle =\left\langle P_{i}\right\rangle ,
\label{m49a} \\
\frac{d}{dt}\left\langle P_{i}\right\rangle &=&\frac{1}{i\hbar }\left\langle %
\left[ P_{i},H_{2BJJ}\right] \right\rangle =-\omega _{i}^{2}\left\langle
Q_{i}\right\rangle ,  \label{m49c}
\end{eqnarray}
which give rise to the following time evolution:
\begin{eqnarray}
\left\langle Q_{i}\right\rangle \left( t\right) &=&\left\langle
Q_{i}\right\rangle \left( 0\right) \cos \omega _{i}t+\frac{\left\langle
P_{i}\right\rangle \left( 0\right) }{\omega _{i}}\sin \omega _{i}t,
\label{m50a} \\
\left\langle P_{i}\right\rangle \left( t\right) &=&\left\langle
P_{i}\right\rangle \left( 0\right) \cos \omega _{i}t-\omega _{i}\left\langle
Q_{i}\right\rangle \left( 0\right) \sin \omega _{i}t.  \label{m50c} \\
\end{eqnarray}
All we need now is to express $J_{x}^{a}$, $%
J_{x}^{b}$ in terms of $Q_{1}$, $Q_{2}$, $P_{1}$, $P_{2}$ by means of Eqs. (%
\ref{m31}), (\ref{m32}), (\ref{m36}), (\ref{m37}); in this way the initial
conditions $\left\langle J_{y}^{a}\right\rangle \left( 0\right) $, $%
\left\langle J_{y}^{b}\right\rangle \left( 0\right) $, $\left\langle
J_{x}^{a}\right\rangle \left( 0\right) $, $\left\langle
J_{x}^{b}\right\rangle \left( 0\right) $ are well known.

Starting from Eqs. (\ref{m36})-(\ref{m37}) we find:
\begin{eqnarray}
Q_{1} &=&\frac{a^{^{\prime }}}{\sqrt{K_{b}}\sqrt{J^{b}}}J_{x}^{b}-\frac{%
b^{^{\prime }}}{\sqrt{K_{a}}\sqrt{J^{a}}}J_{x}^{a},  \label{m51a} \\
Q_{2} &=&\frac{a^{^{\prime \prime }}}{\sqrt{K_{b}}\sqrt{J^{b}}}J_{x}^{b}-%
\frac{b^{^{\prime \prime }}}{\sqrt{K_{a}}\sqrt{J^{a}}}J_{x}^{a},
\label{m51b}
\end{eqnarray}
whose inverse transformation gives $J_x^a$ and $J_x^b$ in terms of $Q_{1,2}$ and permits us to readily obtain the time-evolution of their averages:
\begin{eqnarray}
\left\langle J_{x}^{a}\right\rangle \left( t\right) &=&\frac{a^{^{\prime
}}\left\langle Q_{2}\right\rangle \left( t\right) -a^{^{\prime \prime
}}\left\langle Q_{1}\right\rangle \left( t\right) }{\left[ \frac{a^{^{\prime
\prime }}b^{^{\prime }}}{\sqrt{K_{a}J^{a}}}-\frac{a^{^{\prime }}b^{^{\prime
\prime }}}{\sqrt{K_{b}J^{b}}}\right] }, \\
\left\langle J_{x}^{b}\right\rangle \left( t\right) &=&\frac{b^{^{\prime }}%
\frac{\sqrt{K_{b}J^{b}}}{\sqrt{K_{a}J^{a}}}\left\langle Q_{2}\right\rangle
\left( t\right) -b^{^{\prime \prime }}\left\langle Q_{1}\right\rangle \left(
t\right) }{\left[ \frac{a^{^{\prime \prime }}b^{^{\prime }}}{\sqrt{K_{a}J^{a}%
}}-\frac{a^{^{\prime }}b^{^{\prime \prime }}}{\sqrt{K_{b}J^{b}}}\right] }.
\end{eqnarray}
The coefficients $a^{^{\prime }},b^{^{\prime }},a^{^{\prime \prime }},b^{^{\prime \prime }}$ are defined in the Appendix.
The initial conditions relevant for our study are the one with a small imbalance between the two wells for each species and the coherent initial states.
For the first case we choose $\left\langle J_{x}^{a}\right\rangle \left( 0\right) =\pm 1$, $%
\left\langle J_{x}^{b}\right\rangle \left( 0\right) =\pm 1$, $%
\left\langle J_{y}^{a}\right\rangle \left( 0\right) =0$, $\left\langle
J_{y}^{b}\right\rangle \left( 0\right) =0$, while the particle number is equal to $j_a=j_b=1000.$ Concerning the chosen values of the interaction strengths $\Lambda_a$, $\Lambda_b$ and $\Lambda_{ab}$, in the following we refer to the mixture of $^{85}Rb$ and $^{87}Rb$ atoms realized by the JILA group \cite{exp6}.

\begin{figure}[tbph]
\centering
\includegraphics[scale=0.8]{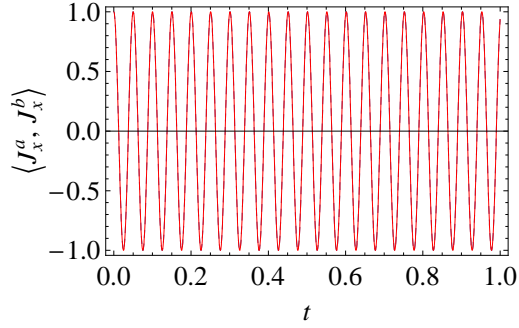}\\
\caption{Behavior of the average value of $J_{x}^{a,b}$ for $\Lambda_a=\Lambda_b$ (units of energy), $\Lambda_{ab}=2.13$, $K_a=K_b=10.$ and initial conditions   $\langle J_{x}^{a}\rangle(0)=\langle J_{x}^{b}\rangle(0)=1.$, $\langle J_{y}^{a,b}\rangle(0)=0$. The time is expressed in units of energy/$\hbar$.}\label{fig:fig1}
\end{figure}

\begin{figure}[tbph]
\centering
\includegraphics[scale=0.8]{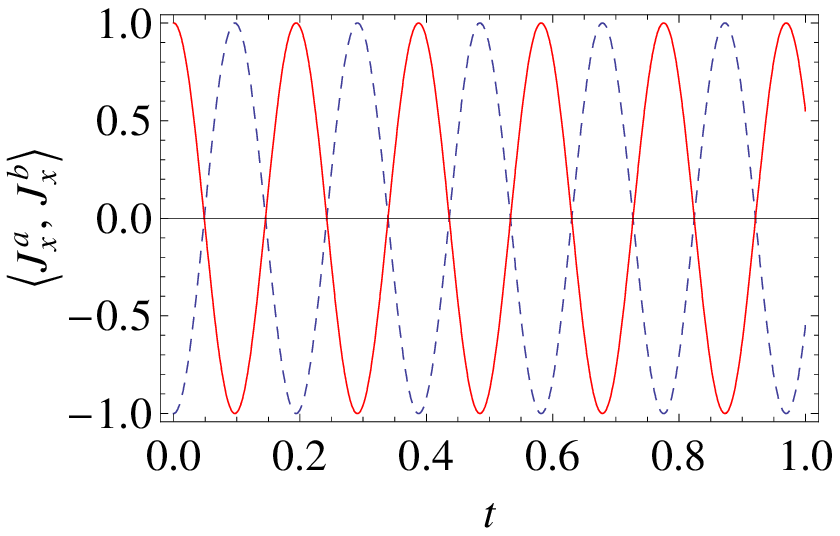}\\
\caption{Behavior of the average value of $J_{x}^{a,b}$ for
$\Lambda_a=\Lambda_b$ (units of energy), $\Lambda_{ab}=2.13$, $K_a=10.,K_b=10.$ and initial conditions
$\langle J_{x}^{a}\rangle(0)=1,\langle J_{x}^{b}\rangle(0)=-1.$, $\langle J_{y}^{a,b}\rangle(0)=0$. The time is expressed in units of energy/$\hbar$.}\label{fig:fig2}
\end{figure}

Figs. \ref{fig:fig1} and \ref{fig:fig2} show the dynamics of $\langle J_x^{a,b}\rangle$ in the case in which there is a small imbalance between the two wells, specifically we consider the case in which there is one unit difference in the left and in the right well, in the absence of imbalance between the two species (the corresponding parameters are reported in the figure caption). Here we note a coherent tunneling between the two wells.

\begin{figure}[tbph]
\centering
\includegraphics[scale=0.8]{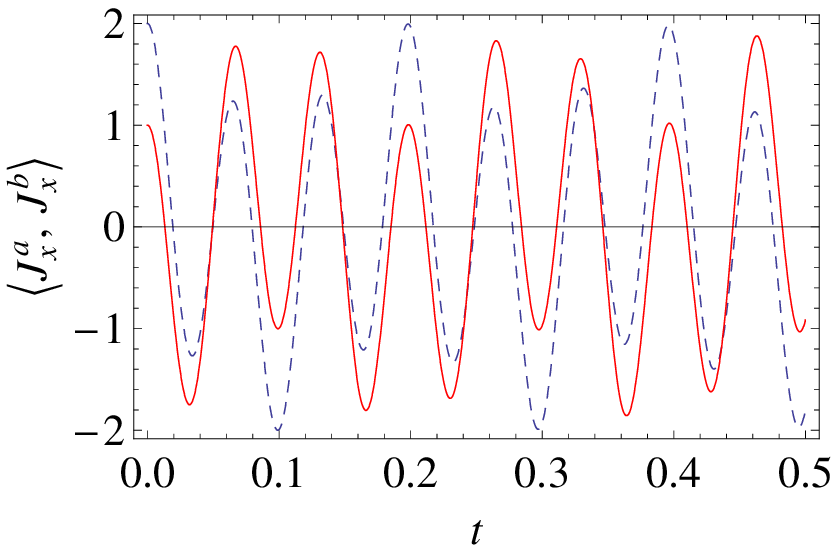}\\
\caption{
Behavior of the average value of $J_{x}^{a,b}$ for
$\Lambda_a=\Lambda_b$ (units of energy), $\Lambda_{ab}=0.8$, $K_a=10.,K_b=10.$ and initial conditions
$\langle J_{x}^{a}\rangle(0)=1.$,$\langle J_{x}^{b}\rangle(0)=2.$, $\langle J_{y}^{a,b}\rangle(0)=0$. The time is expressed in units of energy/$\hbar$.}\label{fig:fig3}
\end{figure}

\begin{figure}[tbph]
\centering
\includegraphics[scale=0.8]{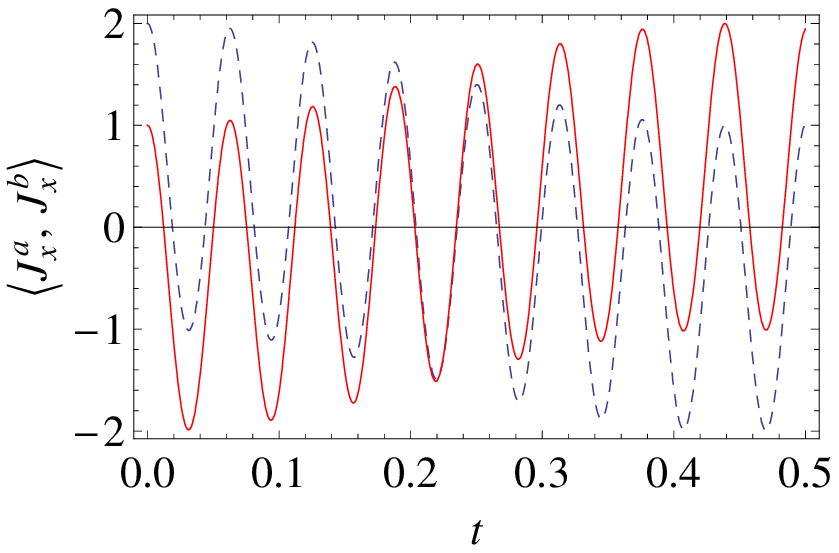}\\
\caption{Behavior of the average value of $J_{x}^{a,b}$ for $\Lambda_a=\Lambda_b$ (units of energy),
 $\Lambda_{ab}=1.$, $K_a=10.,K_b=10.$ and initial conditions   $\langle J_{x}^{a}\rangle(0)=1.$,$\langle J_{x}^{b}\rangle(0)=2.$, $\langle J_{y}^{a,b}\rangle(0)=0$. The time is expressed in units of energy/$\hbar$.} \label{fig:fig4}
\end{figure}

\begin{figure}[tbph]
\centering
\includegraphics[scale=0.8]{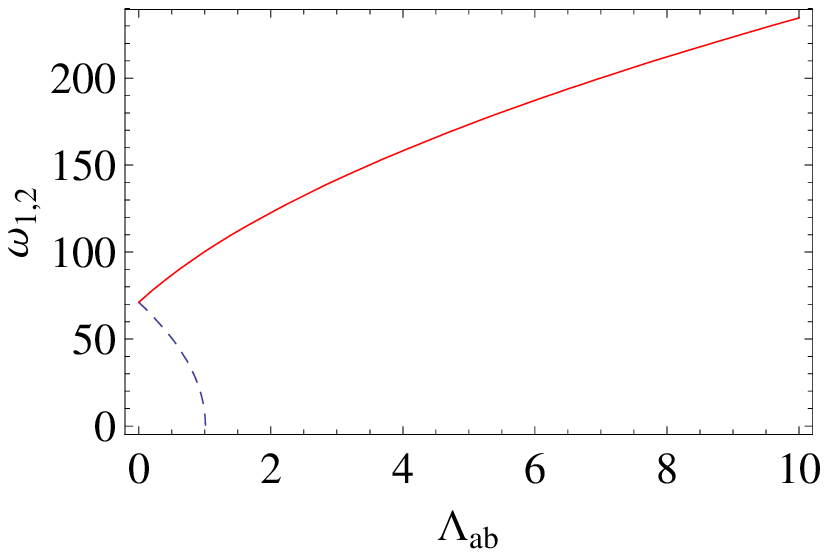}\\
\caption{Behavior of $\omega_1(t)$ (dashed line) and $\omega_2(t)$ (straight line) for $\Lambda_a=\Lambda_b$ (units of energy). }\label{fig:fig6}
\end{figure}

\begin{figure}[tbph]
\centering
\includegraphics[scale=0.8]{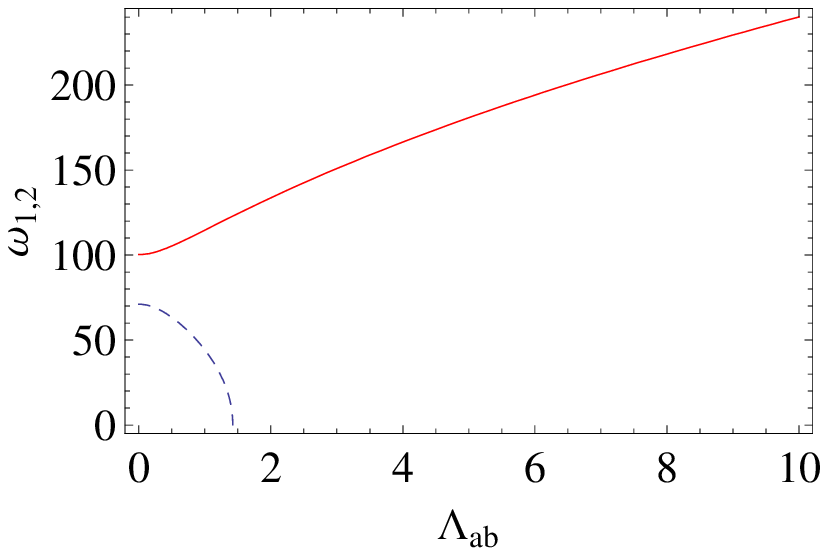}\\
\caption{Behavior of $\omega_1(t)$ (dashed line) and $\omega_2(t)$ (straight line) for $\Lambda_a=2\Lambda_b$ (units of energy).}\label{fig:fig7}
\end{figure}

Figs. \ref{fig:fig3} and \ref{fig:fig4} show instead the behavior of  $\langle J_x^{a,b}\rangle$ in the case of imbalance between the two species, with an imbalance between the two wells of one and two units and for two different values of $\Lambda_{ab}$ (0.8 and 1.). As one can note, at increasing $\Lambda_{ab}$ one approaches a phase separation instability in which the two species tend to separate in the different wells. This behavior can be understood in terms of the behavior of the eigenfrequencies $\omega_{1,2}$ vs $\Lambda_{ab}$. In Fig. \ref{fig:fig6} and Fig. \ref{fig:fig7} one of the two frequency becomes imaginary for a critical value of $\Lambda_{ab}$, thus signalling an instability. Let us note that the instability point is a function of $\Lambda_a,\Lambda_b$ and usually takes place for a critical positive value of the interspecies interaction, as discussed in Section 2, Eqs. (\ref{stab1}) and (\ref{stab2}). In case in which this interaction is attractive the system is always in a coherent tunneling regime.


A few comments on the dynamics of the system are in order here. Compared to our previous analysis\cite{noi1}, the present analysis does not allow the study of long-time scale phenomena since their detection is abruptly increased with $N$, thus only short-time scale effects are reliable. Furthermore we point out that the dynamics should become aperiodic in the general case.

When the initial state is a coherent spin state for each species, $\left|
\psi \left( 0\right) \right\rangle =\left| \psi \left( 0\right)
\right\rangle _{a}\left| \psi \left( 0\right) \right\rangle _{b}$, where $%
\left| \psi \left( 0\right) \right\rangle
_{i}=C_{i}\sum_{m_{i}=-N_{i}/2}^{N_{i}/2}\sqrt{\frac{N_{i}!}{\left( \frac{%
N_{i}}{2}+m_{i}\right) !\left( \frac{N_{i}}{2}-m_{i}\right) !}}\tan
^{m_{i}}\left( \frac{\theta _{i}}{2}\right) e^{-im_{i}\phi _{i}}\left|
m_{i}\right\rangle $, $C_{i}=\sin ^{N_{i}/2}\left( \frac{\theta _{i}}{2}%
\right) \cos ^{N_{i}/2}\left( \frac{\theta _{i}}{2}\right) e^{-i\frac{N_{i}}{%
2}\phi _{i}}$, $i=a,b$, then the initial conditions are:

$\left\langle J_{x}^{a}\right\rangle \left( 0\right) =-\frac{N_{a}}{2}\cos
\theta _{a}$, $\left\langle J_{x}^{b}\right\rangle \left( 0\right) =-\frac{%
N_{b}}{2}\cos \theta _{b}$, $\left\langle J_{y}^{a}\right\rangle \left(
0\right) =\frac{N_{a}}{2}\sin \theta _{a}\sin \phi _{a}$, $\left\langle
J_{y}^{b}\right\rangle \left( 0\right) =\frac{N_{b}}{2}\sin \theta _{b}\sin
\phi _{b}$.

\begin{figure}[tbph]
\centering
\includegraphics[scale=0.8]{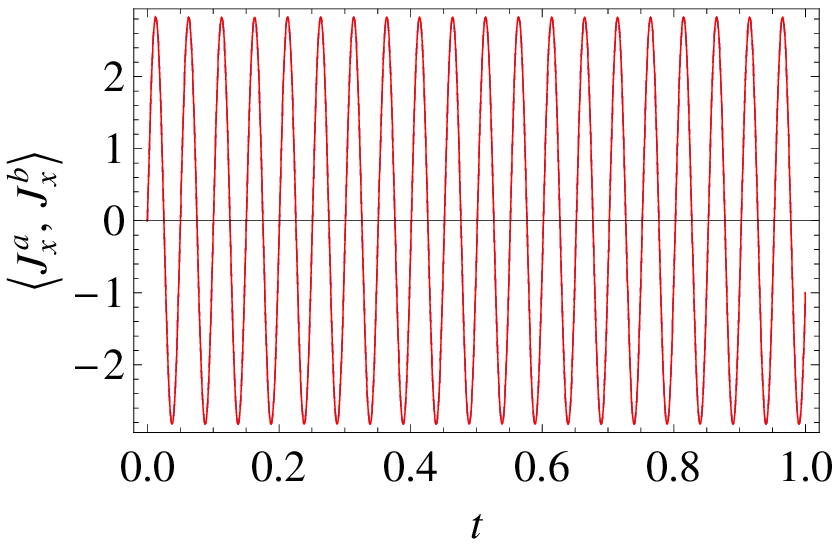}\\
\caption{
Behavior of the average value of $J_{x}^{a,b}$ for $\Lambda_a=\Lambda_b$ (units of energy),
$\Lambda_{ab}=2.13$ and initial conditions with $N_a=N_b=100.$ and $\theta_a=\theta_b=\pi/2.$ and $\phi_a=\phi_b=\pi/4.$. The time is expressed in units of energy/$\hbar$.}\label{fig:fig5}
\end{figure}

In this case the same type of behavior, as for the small imbalance, is observed. In  Fig. \ref{fig:fig5} we take the values $\theta_a=\theta_b=\pi/2.$ and $\phi_a=\phi_b=\pi/4.$

The quantum dynamics above investigated could be experimentally reproduced. If we refer for instance to the mixture of $^{85}Rb$
and $^{87}Rb$ atoms realized by the JILA group \cite{exp6}, a wide
tuning of s-wave interactions is possible via
Feshbach resonances. In particular it is possible to fix the
scattering length of $^{87}Rb$ and to tune the scattering length of $^{85}Rb$ as well as the interspecies one. That allows one to
explore the parameter space in a wide range and to realize
the symmetric regime $\Lambda _{a}=\Lambda _{b}=\Lambda $ as well as the asymmetric one. Furthermore one can tune the inter well coupling, i. e. the parameters $K_{a}, K_{b}$, in such a way to get the semiclassical limit. Another possible realization of the phenomena above described could be obtained with the mixture of $^{41}K$ and $^{87}Rb$ atoms produced by the LENS group \cite{exp5}, which offers a wide possibility of driving from the weak to the strong interacting regime because of the presence of several magnetic Feshbach resonances \cite{simoni}.

\section{Semiclassical dynamics}

In this Section we briefly introduce the semiclassical limit of our model within the linear approximation in order to make a comparison with the quantum results obtained above. A detailed semiclassical analysis has been already carried out in the recent literature (see Refs. \cite{mix1,mix2,mix3,mix4,mix4bis}). Here we only recall the classical equations of motion to give a physical interpretation of $q_{a,b}$ and $p_{a,b}$ in Eq.~(\ref{m31}). From the Hamiltonian (\ref{m27}), we can derive the following equations of motion for the components of the vectors:
$\overrightarrow{J}^{a,b}$:
\begin{eqnarray}
  \label{eq:eom1}
  \frac {dJ^a_x}{dt}&=&-\Lambda_a J_y^a J_z^a - \Lambda_{ab} J_y^a J_z^b,
  \\\label{eq:eom2}
   \frac {dJ^b_x}{dt}&=&-\Lambda_b J_y^b J_z^b - \Lambda_{ab} J_y^b J_z^a,
  \\\label{eq:eom3}
   \frac {dJ^a_y}{dt}&=&\Lambda_a J_x^a J_z^a + \Lambda_{ab} J_x^a
   J_z^b + K_a J_z^a,
  \\\label{eq:eom4}
   \frac {dJ^b_y}{dt}&=&\Lambda_a J_x^b J_z^b + \Lambda_{ab} J_x^b
   J_z^a + K_b J_z^b,
  \\\label{eq:eom5}
  \frac {dJ^a_z}{dt}&=&- K_a J_y^a, \\\label{eq:eom6}
   \frac {dJ^b_z}{dt}&=&- K_b J_b^a.
\end{eqnarray}
These equations imply that
$(\overrightarrow{J}^{a})^2=(J^a_x)^2+(J^a_y)^2+(J^a_z)^2$ and
$(\overrightarrow{J}^{b})^2=(J^b_x)^2+(J^b_y)^2+(J^b_z)^2$ are constants,
so we can introduce:
\begin{eqnarray}
  \label{eq:spherical-a}
 J_x^a=||J^{a}|| \sin \theta_a \cos \varphi_a,
 J_y^a=||J^{a}||\sin \theta_a \sin \varphi_a,
  J_z^a=||J^{a}|| \cos \theta_a,
\end{eqnarray}
and:
\begin{eqnarray}
  \label{eq:spherical-b}
 J_x^b=||J^{b}|| \sin \theta_b \cos \varphi_b,
 J_y^b=||J^{b}||\sin \theta_b \sin \varphi_b,
  J_z^b=||J^{b}|| \cos \theta_b.
\end{eqnarray}

Using (\ref{eq:spherical-a}) and (\ref{eq:spherical-b}) in
(\ref{eq:eom1})-(\ref{eq:eom6}), we obtain the equations\cite{mix3}:

\begin{eqnarray}\label{eom-reduced1}
  \frac {d\theta_a}{dt}&=&K_a \sin \varphi_a, \\\label{eom-reduced2}
   \frac {d\theta_b}{dt}&=&K_b \sin \varphi_b, \\\label{eom-reduced3}
   \frac{d\varphi_a}{dt}&=&(\Lambda_a J^a \cos \theta_a + \Lambda_{ab}
   J^b \cos \theta_b) + K_a \cot \theta_a \cos \varphi_a, \\\label{eom-reduced4}
   \frac{d\varphi_b}{dt}&=&(\Lambda_b J^b \cos \theta_b + \Lambda_{ab}
   J^a \cos \theta_a) + K_b \cot \theta_b \cos \varphi_b.
   \label{eom-reduced}
\end{eqnarray}

These equations coincide with Eqs. (5)-(8) in Ref. \cite{mix3} and Eqs. (5) in Ref.\cite{mix4bis} and Eqs. (3) in Ref.\cite{mix4}.
The energy conservation introduces one extra constraint, so that the
phase space is actually three-dimensional. This may permit in certain
conditions the observation of classical chaos.
If we linearize the Equations (\ref{eom-reduced1})-(\ref{eom-reduced4}) around the point $\theta_a=\theta_b=\pi/2,\varphi_a=\varphi_b=0$, we find the
equations of motion:
\begin{eqnarray}
  \label{eq:eom-linearized1}
  \frac{d \delta \theta_a}{dt} &=& K_a \varphi_a,  \\\label{eq:eom-linearized2}
   \frac {d\delta \theta_b}{dt}&=& K_b  \varphi_b, \\\label{eq:eom-linearized3}
     \frac{d\varphi_a}{dt}&=&-(\Lambda_a J^a \delta \theta_a + \Lambda_{ab}
   J^b \delta \theta_b) -K_a \delta \theta_a, \\\label{eq:eom-linearized4}
     \frac{d\varphi_b}{dt}&=&-(\Lambda_b J^b \delta \theta_b + \Lambda_{ab}
   J^a \delta \theta_a) - K_b \delta \theta_b,
\end{eqnarray}
where $\theta_a=\pi/2+\delta \theta_a$ and $\theta_b=\pi/2+\delta
\theta_b$. These equations derive from the Hamiltonian:
\begin{eqnarray}
  H_{eff}=K_a J^a \frac{\varphi_a^2} 2 +K_b J^b \frac{\varphi_b^2} 2 + \frac
  1 2 \left[ (\Lambda_a (J^a)^2 + K_a J^a) (\delta \theta_a)^2 +
    (\Lambda_b (J^b)^2 + K_b J^b)  (\delta \theta_a)^2 + 2 \Lambda_{ab}
    J^a J^b \delta \theta_a \delta \theta_b\right],
\end{eqnarray}
with the Poisson brackets, $\{\varphi_a,J^a \delta \theta_a\}=1$ and
$\{\varphi_b,J^b \delta \theta_b\}=1$.
By rescaling the variable $\varphi_i$ and $\delta \theta_i$ ($i=a,b$) as $\varphi_i\rightarrow \frac{1}{\sqrt{J^aK_a }}\tilde{\varphi_i}$ and
 $\delta \theta_i\rightarrow \sqrt{J^aK_a }\delta \tilde{\theta_i}$, we do obtain the corresponding classical hamiltonian of (\ref{m33}), with Poisson brackets
$\{\tilde{\varphi_i},J^i \delta \tilde{\theta_i}\}=1$. This Hamiltonian can be diagonalized in a standard way by introducing a linear combination of the variables $\tilde{\varphi_i}$ and $\delta \tilde{\theta_i}$ that preserves the Poisson brackets. The diagonalized Hamiltonian will be that of two independent classical harmonic oscillators of variables $\varphi_1,\varphi_2$ and $\delta\theta_1,\delta\theta_2$.
Applying then the Bohr-Sommerfeld quantization we do reobtain
the spectrum~(\ref{m43}), giving the desired connection between the semiclassical and the quantum approach. This leads also to a physical
interpretation of the conjugate variables $q_{a,b}$ and $p_{a,b}$ in
Eq.~(\ref{m31}) as the azimuthal angles of the pseudospins
$\overrightarrow{J}^{a,b}$. The full classical solution of Eqs. (\ref{eom-reduced1})-(\ref{eom-reduced4}) can be found in Refs. \cite{mix1,mix2,mix3,mix4,mix4bis}.

\section{Conclusions and perspectives}

In this paper we investigated the quantum dynamics of a Bose Josephson junction made of a binary mixture of BECs loaded in a double well potential
within the two-mode approximation. Focussing on the regime where the number of atoms is very large, a mapping onto a $SU(2)$ spin problem together with a Holstein-Primakoff transformation has been performed to calculate the time evolution of the imbalance between the two wells. This approach allows one to exactly solve the system under the assumption of weak interatomic interactions. The results show
an instability towards a phase-separation above a critical positive value of the interspecies interaction while the system evolves towards a coherent tunneling regime for negative interspecies interactions. The detection of a phase separation could be experimentally achieved in current experiments with a mixture of $^{85}Rb$ and $^{87}Rb$ atoms\cite{exp6}.

We point out that all the above results are obtained within the linear approximation. It would be interesting to extend our model beyond the linear regime; in such a case the classical dynamics may exhibit a chaotic behavior in some parameter range because the phase-space is three dimensional. At the quantum level, these features will show up in the spectrum as well as the eigenstates of the Hamiltonian. Indeed the Hamiltonian is not time reversal invariant because of the terms linear in $J_x/J_z$, and we expect that the distribution of spacings between energy levels should follow the GUE (Gaussian Unitary ensemble) statistics
\cite{mehta}. Regarding the dynamics, we conjecture that the short time scale behavior of the quantum system will look chaotic, but the long time behavior will not. Such an analysis will be carried out in detail in a forthcoming publication.


\appendix

\section{Coefficients}
The coefficients $a^{^{\prime }}$, $b^{^{\prime }}$, $%
a^{^{^{\prime \prime }}}$, $b^{^{\prime \prime }}$ are defined as follows:
\begin{eqnarray}
a^{^{\prime }} &=&\frac{2\omega _{ab}}{\sqrt{4\omega _{ab}^{2}+\left[ \left(
\omega _{b}^{2}-\omega _{a}^{2}\right) +\sqrt{\left( \omega _{a}^{2}-\omega
_{b}^{2}\right) ^{2}+4\omega _{ab}^{2}}\right] ^{2}}},  \label{m53a} \\
b^{^{\prime }} &=&\frac{\left( \omega _{b}^{2}-\omega _{a}^{2}\right) +\sqrt{%
\left( \omega _{a}^{2}-\omega _{b}^{2}\right) ^{2}+4\omega _{ab}^{2}}}{\sqrt{%
4\omega _{ab}^{2}+\left[ \left( \omega _{b}^{2}-\omega _{a}^{2}\right) +%
\sqrt{\left( \omega _{a}^{2}-\omega _{b}^{2}\right) ^{2}+4\omega _{ab}^{2}}%
\right] ^{2}}},  \label{m53b} \\
a^{^{\prime \prime }} &=&\frac{2\omega _{ab}}{\sqrt{4\omega _{ab}^{2}+\left[
\left( \omega _{b}^{2}-\omega _{a}^{2}\right) -\sqrt{\left( \omega
_{a}^{2}-\omega _{b}^{2}\right) ^{2}+4\omega _{ab}^{2}}\right] ^{2}}},
\label{m53c} \\
b^{^{\prime \prime }} &=&\frac{\left( \omega _{b}^{2}-\omega _{a}^{2}\right)
-\sqrt{\left( \omega _{a}^{2}-\omega _{b}^{2}\right) ^{2}+4\omega _{ab}^{2}}%
}{\sqrt{4\omega _{ab}^{2}+\left[ \left( \omega _{b}^{2}-\omega
_{a}^{2}\right) -\sqrt{\left( \omega _{a}^{2}-\omega _{b}^{2}\right)
^{2}+4\omega _{ab}^{2}}\right] ^{2}}}.  \label{m53d}
\end{eqnarray}

\newpage

\end{document}